\RequirePackage{fix-cm}
\documentclass{svjour3}                     
\smartqed  
\long\def\/*#1*/{}
\usepackage{graphicx}
\usepackage{amssymb, amsmath}
\usepackage{color}
\usepackage{cite}
\usepackage[numbers]{natbib}
\usepackage{ulem}

\definecolor{zeg} {RGB}{255,0,120}

\definecolor{orange} {RGB}{255,120,0}

\begin{document}

\title{Route to chaos in a coupled microresonator system with gain and loss}

\author{Krzysztof Zegadlo, Nguyen Viet Hung$\dagger$, Vladimir V. Konotop, Jakub Zakrzewski and  Marek Trippenbach}

\authorrunning{K. Zegadlo et. al.} 
\titlerunning{Chaos in  coupled microresonators}

\institute{Krzysztof Zegadlo and Jakub Zakrzewski\at
              Instytut Fizyki im. Mariana Smoluchowskiego,Uniwersytet
Jagiello\'{n}ski, Lojasiewicza 11, PL--30--348 Krak\'{o}w, Poland,\\
           Nguyen Viet Hung \at
              Advanced Institute for Science and Technology, Hanoi
University of Science and Technology, Hanoi, Vietnam,  \\
Marek Trippenbach \at Faculty of Physics,
University of Warsaw, ul. Pasteura 5, PL--02--093 Warszawa, Poland. \\
Vladimir V. Konotop\at Centro de F\'isica Te\'orica e Computacional
and Departamento de F\'isica, Faculdade de Ci\^encias, Universidade
de Lisboa, Campo Grande, Edif\'icio C8, Lisboa 1749-016,
Portugal.\\
$\dagger$Correspondence should be addressed to:\\
Email: hungvn1102@gmail.com; hung.nguyenviet1@hust.edu.vn}

\date{Received: date / Accepted: date}

\maketitle

\begin{abstract}
We consider chaotic dynamics of a system of two coupled ring
{resonators} with a linear gain and a nonlinear absorption. Such a
structure can be implemented in various settings including
microresonator nanostructures, polariton condensates, optical
waveguides or atomic Bose-Einstein condensates of ultra-cold atoms
placed in a circular-shaped trap. From the theoretical point of view
this system is attractive due to its modulational instability and
rich structure, including various types of spontaneous symmetry
breaking, period doubling bifurcations, eventually leading to
chaotic regime. It is described by set of partial differential
equations but we show that the so called Galerkin approximation can
explain most of the system characteristics mapping it on the
dynamics of few coupled oscillator modes. The main goal of present
study is to investigate various routes to chaos in our non-hermitian
system  and to show the correspondence between the continuous
operator problem and its discrete representation.


\end{abstract}

\section{Introduction}
\label{intro} While it is quite common to consider physical models
with a linear loss, fibre optics serves as one of the areas where
nonlinear loss phenomena, due to stimulated Brillouin or Raman
scattering \cite{Smith72,Agra01,Boyd} play an important role. These
effects are effectively  described by Kerr-like (cubic)
nonlinearities. Moreover, this nonlinearity bridges the two
seemingly remote areas of physics such as the nonlinear optics and
cold atom physics. It was revealed about 20 years ago, while
considering wave mixing in atomic condensates \cite{4WM,atomlaser}.
{For atomic condensate the loss may be due to two or three body
inelastic interactions}. Note also that strong nonlinear loss is
present in Bose-Einstein condensate (BEC) of exciton polaritons in a
semiconductor microcavity \cite{dakis,newbook}, which is a
macroscopically populated coherent quantum state subject to
concurrent pumping and decay.

On that account the model that we propose to study here can be
applied to both BECs of exciton polaritons and to optical
nanostructures of coupled microresonators. One and the other belong
to the rapidly developing fields of both theoretical and
experimental physics, receiving lately a lot of attention. The
former consists of bosonic quasi-particles that exist inside
semiconductor microcavities, in the form of a superposition of an
exciton and a cavity photon and exhibit a transition to quantum
degeneracy akin to (BEC) \cite{EPC1,EPC2,EPC3,EPC4,EPC5}. The latter
holds the possibility to create a new platform for biological
sensing applications~\cite{plat}. Optical microresonators were also
investigated in the context of switches~\cite{switches1,switches2},
lasers,~\cite{lasers1,lasers2,lasers3}, temporal solitons
\cite{optsol1,optsol2,optsol3,optsol4}, soliton frequency combs
\cite{freqcomb1,freqcomb2,freqcomb3}, and even {a} transition to
chaos \cite{michaos}.

In the following we shall consider a particular example of such a
system - a two coupled rings structure. This geometry has numerous
interesting features from both, theoretical (periodic boundary
conditions, indicating mode quantization) and experimental (easy
fabrication in nano-structured systems, potential applications)
perspectives.

The model is presented in Sec.~\ref{model}. It was studied by some
of us earlier revealing rich dynamical behavior such as the
spontaneous symmetry breaking, modulational instability or complex
and persistent flow of currents between the rings
\cite{SciRep,Symmetry}.  In the present work we concentrate on the
route of chaos, which may be observed in this system as demonstrated
numerically in Sec.~\ref{route}.

Previously we demonstrated {that} in the regime of regular motion
the system is well described by an analytic few mode Galerkin
approximation (GA) \cite{Ern} {originally proposed for solving of
the Laplace equation in the electric field \cite{galerkin}}. In
mathematics, in the area of numerical analysis, Galerkin methods are
a class of methods for converting a continuous operator problem
(such as a differential equation) to a discrete problem. It uses a
linear combination of basis functions with weights determined by
minimizing a globally integrated measure of the error in the
solution. It is widely used in the analysis of various scientific
and engineer problems (see for example \cite{Chekroun,Wanga})
{including models described by the Gross-Pitaevskii equation (GPE)
with harmonic oscillator potential \cite{GA-Malomed} or in
multi-component space \cite{Driben16,Wang17}. There is a broad
spectrum of GA applications in descriptions of physical and chemical
processes, where statistical models are used \cite{apps1,apps2} and
the Galerkin expansion allows to significantly simplify the problem.
Here we prove that this approach can also help to understand the
chaotic instability of our system, and when appropriate number of
modes is included, is practically indistinguishable from the
continuous model.

\section{The model}
\label{model} We consider two coupled rings structure. Apart from a
cubic nonlinearity,  we assume that the system is interacting with the
environment. This interaction is characterized by the presence of a
linear gain and nonlinear losses. We adopt the mean field model
\cite{SciRep} described by a pair of Gross-Pitaevskii equations
(GPE) coupled by a linear homogeneous term, characterized by a
constant coupling $c$
\begin{eqnarray}
i\frac{\partial \Psi_1}{\partial t}=-\frac{\partial^2 \Psi_1}{\partial x^2}+i\gamma \Psi_1 +\left(1-i\Gamma\right)\vert \Psi_1 \vert^2 \Psi_1+c\Psi_2 \notag \\
i\frac{\partial \Psi_2}{\partial t}=-\frac{\partial^2
\Psi_2}{\partial x^2}+i\gamma \Psi_2 +\left(1-i\Gamma\right)\vert
\Psi_2 \vert^2 \Psi_2 + c\Psi_1. \label{GPE system}
\end{eqnarray}
Here the term with $\gamma$ is responsible for a linear gain, while
that with $\Gamma$ quantifies nonlinear losses (all parameters are
assumed to be homogeneous across the whole structure). We choose the
units in such a way that interactions within each {ring} lead to a
Kerr-like nonlinearity of unit strength. The assumed geometry
implies that $x\in\left\langle 0,2\pi\right)$ is an angular variable
and periodic boundary conditions are applied, i.e. $\Psi_i
\left(x,t\right)=\Psi_i \left(x+2\pi,t\right)$. Due to these
boundary conditions the system reveals an apparent quantization with
only discrete modes being allowed for the dynamics. Note that the
same set of equations \eqref{GPE system} describes a single ring
with a spinor (spin=1/2) condensate with spin components coupled by
$c$-proportional term, but also the exciton-polariton systems in the
microcavities and even nano-photonics systems \cite{nano}. The
system (\ref{GPE system}) appears also as a simplified model with
{a} quintic nonlinearity and diffusion introduced and studied
numerically subject to zero boundary conditions in \cite{Malomed}.

The system described by Eq.~(\ref{GPE system}) reveals very rich
dynamics \cite{SciRep} and is very sensitive to both the change of
the initial conditions and the particular values of parameters, in
particular,  to the value of the coupling between rings, {denoted by} $c$. One may
find different (symmetric and antisymmetric) solutions of \eqref{GPE
system}, as it is discussed at lengths in our previous study
\cite{SciRep}. Interestingly stationary inhomogeneous states also
exist - they manifest spontaneous translational symmetry breaking.
Following \cite{SciRep} let us denote by $\kappa$ the following
integral
\begin{equation}
\kappa=\frac{1}{2\pi}\int_0^{2\pi} \frac{\partial}{\partial x}
{\mathrm arg} (\Psi_i) dx. \label{topolo}
\end{equation}
Due to the ring quantization $\kappa$ takes integer values only and will
be referred to as a topological charge in the spirit of a vortex
terminology.

One can check that during the time evolution the topological charge
does not have to be conserved, so it is relatively easy to observe
stable vortices. Indeed our coupled system permits solutions of the
form (out-of-phase)
\begin{eqnarray}
\label{vortex1}
\Psi_1(x,t)=-\Psi_2(x,t)=\sqrt{\frac{\gamma}{\Gamma}}
e^{i\left[\kappa x-(\frac{\gamma}{\Gamma}+\kappa^2-c)t\right]}
\end{eqnarray}
which are stable for the whole range of parameters $\Gamma, \gamma$
and $c$, in analogy with antisymmetric solutions. There exists also
 in-phase solutions
\begin{equation}
\Psi_1(x,t)=\Psi_2(x,t)=\sqrt{\frac{\gamma}{\Gamma}}e^{i\left[\kappa
x-(\frac{\gamma}{\Gamma}+\kappa^2+c)t\right]}. \label{vortex2}
\end{equation}
All modes mentioned above are stable for specific ranges of
parameters and in \cite{SciRep} we constructed a two dimensional map
in $\gamma/\Gamma$ and $c$ to characterize the spectrum of stable
solutions. In \cite{SciRep} we also discussed stability condition,
applying linear stability analysis. In the present study we focus on
possible instabilities describing in detail the "route to chaos"
scenarios. The results are presented in the next section, followed
by an approximate analysis based on the Galerkin approximation.

\section{Route to chaos}
\label{route} We consider the evolution described by Eq.~(\ref{GPE
system}) for given values of control parameters: the gain $\gamma$,
the  loss $\Gamma$ and the coupling between rings, $c$. To address
the issue of a transition from a regular to a chaotic behavior we
sample the dynamics {in various configurations}. First we keep the
gain and loss parameters at constant values ($\Gamma=1$ and
$\gamma=1.75$) and we compare the solutions for different values of
the coupling $c$. Then we keep $c$ and $\gamma$ fixed and vary
$\Gamma$, and finally we present two dimensional $c-\gamma$ map for
$\Gamma=1$. In each case we start from the same kind of the initial
conditions, slightly perturbed unstable, symmetric, stationary state
\begin{equation}
\Psi_{1,2}\left(x,0\right)=\sqrt{\frac{\gamma}{\Gamma}}\left[1 \pm
\sum_{k=1}^{l} A^k_0 cos\left(k x\right)\right], \label{numerical input}
\end{equation}
with upper and lower signs corresponding to the first and second
ring, respectively. The small perturbation amplitude is equal to
$A^k_0=10^{-2}$ for all $k$. Notice that due to periodic boundary
conditions we have a discrete spectrum of available spatial modes
$k$. For a regular regime, a choice of $A_0^k$ seems not important.
We have checked  that long time evolution is not affected by the
choice of an initial perturbation. However, as we shall see on the
example below even a small modification of the initial conditions
affects the long time solution obtained in the "chaotic" regime.

We have observed that asymptotically, for each set of control
parameters, there is a clear behavior that can be classified
within the limited set of categories: an antisymmetric state, a
vortex state (as defined above), a limit cycle (with one or more
frequencies), or finally, an irregular motion that reveals the
features of the chaotic dynamics. In the previous study
\cite{SciRep} we presented in a compact way the whole gallery of
these states in some range of parameters. Here we focus on a small
range of states to show possible "routes to chaos".

\begin{figure*}[th]
\centering
\includegraphics[scale=0.35]{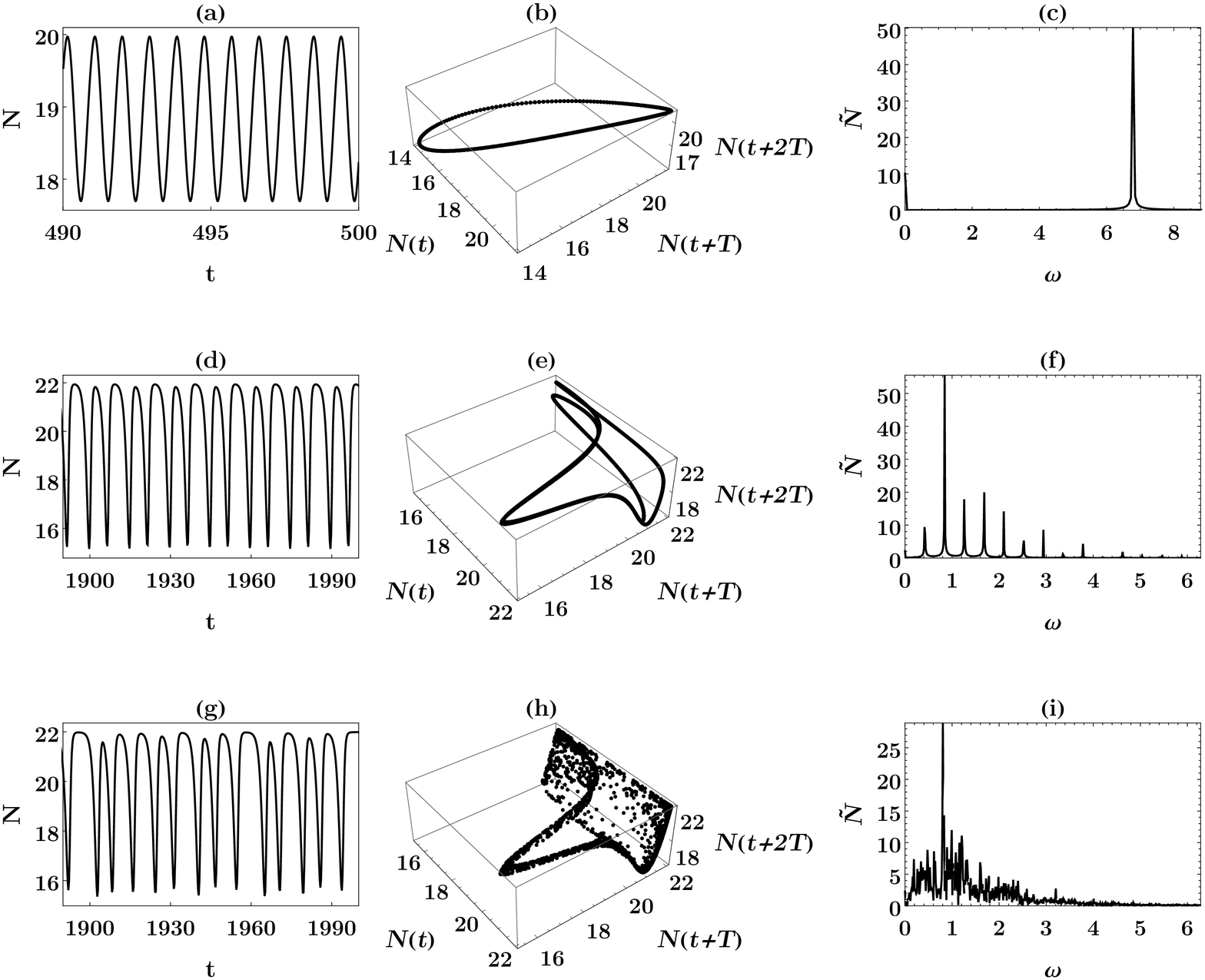}
\caption{(a,d,g) Total norm \eqref{norm} as a function of time is
shown in the left column, panels (a,d,g). The middle column, panels
(b,e,h) show the trajectories  in the time-delay diagram (sometimes
called phase-portrait), using total norm as a variable.  The right
column (c,f,i) presents $\tilde{N}(\omega)$ modulus of Fourier
transform of the total norm (\ref{norm}). The
results correspond to the limit cycle solution with one frequency
(with $c=1.74$ - upper row), many frequencies ($c\approx 1.8$ -
middle row)  and for the chaotic behavior ($c=1.81$ - bottom row).}
\label{behaviour plot}
\end{figure*}

Several observables may be defined to characterize our system, some
of them quite complex. We shall look {at first} at one of the
simplest. This quantity, which is used by us to characterize
different kinds of possible states is a total norm defined as
\begin{equation}
N\left(t\right)=\int_{0}^{2\pi} \left[
\vert\Psi_1\left(x,t\right)\vert^2 +
\vert\Psi_2\left(x,t\right)\vert^2 \right] \,dx. \label{norm}
\end{equation}
Such a  norm is preserved in Hermitian systems. Our
system is non-conservative and in general the total norm as well as
the total energy, and even the topological charge are not
necessarily conserved.

Fig.~\ref{behaviour plot} represents graphically typical situations
found for a long time dynamics of the system. In the regular regime,
say, at $c=1.72$ to $c=1.73$, we observe a stationary behavior with
the norm constant in time. In this case it is an antisymmetric
state. Upon increasing the coupling $c$ we enter the regime of
non-stationary states. For a relatively low coupling we first
observe simple regular oscillations (cf. Fig~\ref{behaviour plot}a).
To extract more information about the system in the non-stationary
regime we generate, for particular (characteristic) points on the
contour profile plot, a set of figures, including, besides the time evolution of
the total norm, also a spatial Fourier transform at the output of the
time evolution, supplemented with a time delayed phase portrait. The
latter is based on Takens-Mane (TM) theorem \cite{takens,mane} which
allows us for a projection of the one-dimensional variable onto a
multidimensional space (in our case 3D). The dynamics of a chaotic
{system} takes place in some real space which is unavailable to our
investigation. The total norm which is here the single observable
reflects the one-dimensional projection of the true
multi-dimensional evolution. To avoid any artificial crossings of
the orbit (caused by the projection -- not present in the real
dynamics) the so called unfolding is applied according to the TM
theorem. More independent variables than one are obtained by using a
single scalar variable (in our case the total norm, defined at
\eqref{norm}) calculated at three different instants of time: $t$,
$t+T$ and $t+2T$, where  $T$ is a predefined delay of time.
The excellent discussion of correlations among data points,
proper choice of time delay, appropriate number of coordinates and the
relation to the average mutual information can be found in \cite{abarbanel}.

The phase portraits obtained in this way are presented in the middle
column in Fig.~\ref{behaviour plot}. Note that such an unfolding of
a single quantity helps in recognizing the chaotic signatures of
motion which are difficult to identify otherwise in the time
dynamics. Periodic oscillations of the total norm from
Fig.~\ref{behaviour plot}a are represented by  the solid close
circuit trajectory in Fig.~\ref{behaviour plot}b. To describe it
quantitatively, the Fourier transform of the total norm with respect
to time is added in the right panel of Fig.~\ref{behaviour plot}c.
The spectrum contains a single frequency which corresponds to the
period of the norm oscillations. This type of motion can be
classified as a non-stationary regular motion with a single period.
Upon increasing the coupling $c$, the dynamics of the system becomes
more complicated. Oscillations remain regular but are no longer
characterized by a single period. Properties of that case are visible
 in the middle row of Fig.\ref{behaviour plot} (cf.
Fig.~\ref{behaviour plot}d) for the time evolution of the norm and
the system trajectory in Fig.~\ref{behaviour plot}e). That behavior
is similar to the case presented in the top row but exhibits several
oscillations with different periods and is quite typical for the
transition towards a fully chaotic motion. Both cases presented in
the top and the middle row belong to the class of limit cycles and
the system goes from the single frequency behavior (top) to the
multi-frequency case (according to the Fig.\ref{behaviour plot}f) by
the period doubling scenario.

Finally, there exist the limiting value of the coupling parameter
where a transition into chaos appears. This is a point where a
regular periodic motion transforms into an irregular one, as
presented in Fig.~\ref{behaviour plot}g. Then the solution in the
phase portrait (compare Fig.~\ref{behaviour plot}h)) starts to fill
a bounded region of the phase space. {The} multi-frequency solution
transforms into the one with a continuous Fourier spectrum as shown
in Fig.~\ref{behaviour plot}i. The bounded region of dynamics in the
phase space results from the volume contraction (typical in
dissipative systems) and drives the system to a null-volume chaotic
attractor. This is apparent in Fig.\ref{behaviour plot}h where phase
points are irregularly distributed but the region of their existence
is contained in the 2-dimensional manifold. The chaotic system is
very sensitive to initial conditions as expected. Even a small
perturbation will provide a completely different trajectory.

\begin{figure}[th]
\centering
\includegraphics[scale=0.4]{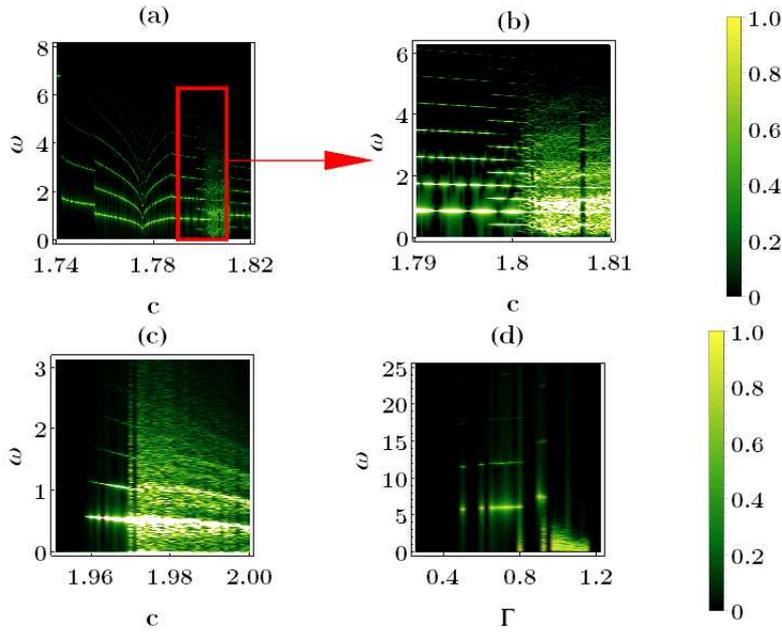}
\caption{Route to chaos Bifurcation diagram. In panel (a): modulus
of the Fourier transform of the total norm, $N(t)$, (see
Eq.(\ref{norm})) for the solution of Eqs.(\ref{GPE system}), for
$\Gamma=1$, $\gamma=1.75$ and $c$ in the range of $\left[1.74,
1.82\right]$. In panel (b): magnification of the region enclosed by
the red rectangle. Notice the leading frequency component at $\omega
\approx 1$ and a narrow stripe around $c \approx 1.808$, where the
motion becomes suddenly regular. Panel (c) displays a bifurcation
diagram for another route (corresponding to $\Gamma=1$ and
$\gamma=2$) versus the control parameter $c$, in the range
$\left[1.95, 2.0\right]$. In panel (d) the diagram is plotted for
fixed $\gamma=2$ and $c=1.98$ for varying $\Gamma\in \left[0.25,
1.225\right]$.} \label{bifurcation diagram}
\end{figure}

The general conclusion that we derive from our necessarily limited
studies is that the change of the long time (asymptotic) behavior is
rather dramatic and sensitive to the values of parameters. A small
change of the chosen control parameter ($c$, $\gamma$ or $\Gamma$)
leads to the transition from a regular asymptotic state to a chaotic
motion, via a process resembling a standard period doubling
scenario, known from classical dynamics of simple low dimensional
systems \cite{schuster,strogatz}.

We demonstrate such a dramatic transition in Fig.~\ref{bifurcation
diagram}a revealing an interesting structure of frequencies in the
limit cycle (LC) regime. We highlight the dominant peaks in the
Fourier transform of the norm, Eq. \eqref{norm},  as the function of
the coupling $c$. The analysis has been done for the range of
coupling parameter $c\in \left(1.73, 1.83\right]$ (here $\Gamma=1$
and $\gamma=1.75$). For the value of $c$ below the lower bound the
system admits regular solutions only. They can be antisymmetric or
asymmetric with a constant homogeneous amplitude or even  can be
inhomogeneous with respect to the spatial coordinate $x$ thus
revealing {a} translational symmetry breaking \cite{SciRep}. All
these solutions are treated within this paper as regular fixed
points which are sinks (attractors). We observe regular stationary
solutions that are constant in time and the Fourier transform (FT)
contains only one central peak with zero frequency. Then the first
symmetry breaking occurs and {the} FT shows single peaks
corresponding to the limit cycle state with a single period. For a
still bigger $c$ more frequencies appear, still, however, in the
limit cycle regime. For $c\approx 1.76$ there is the Hopf
bifurcation where {a} new stable set of frequencies appears at the
expense of the former one, which is no longer stable. For bigger
values of the control parameter, closer to the chaotic regime, we
observe a typical period doubling scenario. At the bifurcation point
each branch of the diagram becomes unstable yielding two or more new
branches emerging from it. That is the sub-critical bifurcation
(phase transition of the first kind), i.e. there is a hysteresis,
but  states from the backward going branches cannot be observed
being unstable. For the $c$ value just above 1.8 we observe a
typical chaotic behavior. The spectrum of the frequencies is now
smeared over a wide region, ranging from zero up to about 4 (see
also Fig.~\ref{spectra}a). Additionally one can distinguish bright
bands corresponding to leading frequencies. Hence there is still
some regularity in the chaotic region. We observe here a certain
similarity with the so called exploding solitons introduced by
Akhmediev \cite{Akhmediev1,Akhmediev2,Akhmediev3}.

Another interesting feature present in the chaotic regime is a very
thin stripe with no chaos. Moreover, for $c$ values bigger than
those presented in Figs.\ref{bifurcation diagram}a (and enlarged in
Fig. \ref{bifurcation diagram}b) chaos suddenly disappears. Such a
behavior is referred to as a "crisis" and is a rather common
property of chaotic systems. The points of transition from chaos to
stable frequencies are typical saddle-node bifurcations, where
single stable branches emerge simultaneously with the unstable ones
(cf. chaotic scenarios in \cite{schuster}).

As a second case we have chosen a path in the different part of the
parameter space, with fixed gain and loss amplitudes ($\Gamma=1$ and
$\gamma=2$) and the coupling $c$ within the range from $c=1.95$ to
$c=2$. The result is shown in Figure~\ref{bifurcation diagram}c
again as a plot of FT of the total norm, \eqref{norm},  for
asymptotic solutions corresponding to {different} fixed vales of
$c$. Here we observe how the system is approaching chaotic regime,
evolving from regular solutions for lower values of $c$ and
characterized by a constant norm, entering chaotic region via one
subcritical bifurcation. On the other side of the chaotic regime, we
observe again an inverse saddle-node bifurcation, exactly as in the
previous case.

For a sake of comparison we have investigated another possibilities
too, when the coupling value $c$ is fixed and the route to chaos
depends either on the gain $\gamma$ or on the loss $\Gamma$ as a
variable. The latter scenario is plotted in Fig.~\ref{bifurcation
diagram}d and confirms former conclusions formulated for {the}
changing $c$ case. A scenario with changing $\gamma$ is not plotted
here but is included in the general map of solutions ($c$ vs
$\gamma)$ in the next section.

Transition to chaos may be observed in other dynamical parameters.
To exemplify this idea we will refer to the currents mentioned
briefly above and introduced in our previous study \cite{SciRep}. We
define current densities along each ring
\begin{eqnarray}
\label{current_long}
j_{\alpha}(x,t)=\frac{1}{2i}\left(\Psi_\alpha^*\frac{\partial \Psi_\alpha}{\partial x}-\Psi_\alpha\frac{\partial \Psi_\alpha^*}{\partial x}\right), \quad \alpha=1,2
\end{eqnarray}
and the current density between the two rings
\begin{eqnarray}
\label{current_ort}
j_\bot(x,t)=\frac{c}{2i}\left(\Psi_1^*\Psi_2-\Psi_1\Psi_2^*\right).
\end{eqnarray}
For each of these current densities one can define total current by
integrating density over the length of the ring. Total currents are
obviously functions of time and can be analyzed in the identical
manner as we did with the total norm. In particular one can plot
Fourier transform of currents for different values of the coupling
coefficient. Fig.~\ref{currents} shows that bifurcation diagrams can
be obtained for different dynamical characteristics, if they are
non-trivial. For instance in the case illustrated in
Fig.~\ref{currents}(a) the current between rings in vanishing and we
observe two identical and counter-propagating currents in each ring.
When transformed to the Fourier space they may be used as chaos
indicator instead of the total norm. In the case shown in
Fig.~\ref{currents}(b) and Fig.~\ref{currents}(c) both currents
exist and they contain the same information as far as the onset of
chaos is concerned.

\begin{figure}[th]
\centering
\includegraphics[scale=0.36]{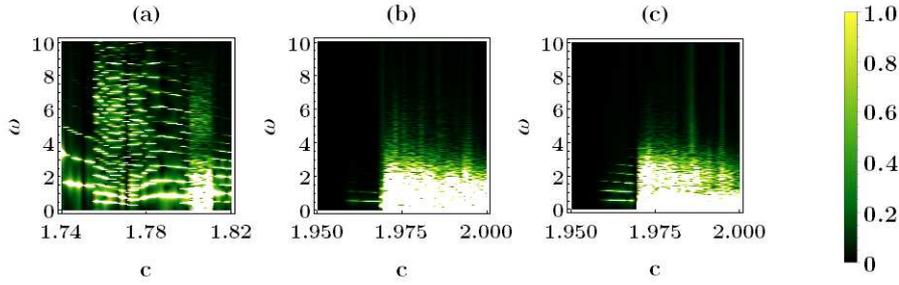}
\caption{Bifurcation diagram of the route to chaos obtained from the total current.
In panel (a) we plot Fourier transform of the current in each ring corresponding to Fig.~\ref{bifurcation diagram}(a), for $\Gamma=1$ and $\gamma=1.75$. Panels (b) and (c) correspond to Fig.~\ref{bifurcation diagram}(c) with  $\Gamma=1$ and $\gamma=2$ and present Fourier transforms of the total current in one of the rings and between rings, respectively.}
\label{currents}
\end{figure}

\section{Galerkin approximation. Few modes model.}
\label{galerkin}

The number of explicitly given analytical solutions of \eqref{GPE
system} covers only a small fraction of available possibilities. To
capture the essence of the transition from a regular behavior to the
chaotic regime we apply the so called Galerkin approximation (GA)
\cite{SciRep,Ern,galerkin,Chekroun,Wanga,GA-Malomed,Driben16,Wang17}.
The idea of GA is based on the observation that due to the periodic
boundary conditions only certain $k$ vectors participate in the
dynamics. It turns out that full continuous dynamics can be well
approximated and viewed as the interaction of (small) number of
modes, even in the chaotic regime. The computational problem is then
reduced to a solution of few ordinary differential equations instead
of considering the full model described numerically by partial
differential equations. That allows us to save the computational
time and the results can be obtained very efficiently. The
wavefunction in each ring may be expressed in terms of Fourier
series

\begin{equation}
\psi_\alpha \left(x,t\right)=\sum_{n=-\infty}^{\infty}B_{\alpha,n}\left(t\right)e^{\mathrm{i}n x}, \label{Fourier expansion}
\end{equation}
where $\alpha={1,2}$ denotes the rings. While in
\cite{SciRep} these solutions were expressed in the stationary form
with all amplitudes $B_{\alpha,n}$ independent of time, in general,
they can be time dependent. By substituting \eqref{Fourier
expansion} into \eqref{GPE system} one obtains

\begin{eqnarray}
\mathrm{i}\frac{\mathrm{d} B_{1,n}}{d
t}=\left(\mathrm{i}\gamma+n^2\right)B_{1,n}+\left(1-\mathrm{i}\Gamma\right)\sum_{k=-l}^{l}\sum_{m=-l}^{l}B_{1,k+m-n}^{\ast}B_{1,k}B_{1,m}+c B_{2,n} \notag \\
\mathrm{i}\frac{\mathrm{d}
B_{2,n}}{dt}=\left(\mathrm{i}\gamma+n^2\right)B_{2,n}+\left(1-\mathrm{i}\Gamma\right)\sum_{k=-l}^{l}\sum_{m=-l}^{l}B_{2,k+m-n}^{\ast}B_{2,k}B_{2,m}+c
B_{1,n} \label{Fourier components equations}
\end{eqnarray}
where $2l+1$ is the number of modes taken into account. It turns out
that restricting the dynamics to just a few modes (typically we take
$l=5$) is sufficient. The GA gives an excellent agreement with the
full system predictions. In particular the Fourier spectra in the
case of a limit cycle behavior, both for the single frequency and
the multi-frequency cases [shown, respectively, in
Fig.~\ref{behaviour plot}c and Fig.~\ref{behaviour plot}f],
calculated solving GPE (\ref{GPE system}) and using GA are
practically indistinguishable. It is directly illustrated in
Fig.~\ref{spectra}a and Fig.~\ref{spectra}b, where a side-by side
comparison of  dynamics in the time domain is plotted. In addition
Fourier spectra of both full dynamics ($\tilde{N}_1$) and GA
($\tilde{N}_2$) may be determined, again demonstrating close
similarity as shown in Fig.~\ref{spectra}c and in the lower row of
panels, where we present the relative error defined as
$\Delta\tilde{N}/\tilde{N}_{tot}=(\tilde{N}_1-\tilde{N}_2)/(\tilde{N}_1+\tilde{N}_2)$.
More importantly, the bifurcation diagrams (as those presented in
Fig.~\ref{bifurcation diagram}) obtained using GPE solutions and the
GA agree fully within our resolution accuracy.

\begin{figure*}[th]
\centering
\includegraphics[scale=0.41]{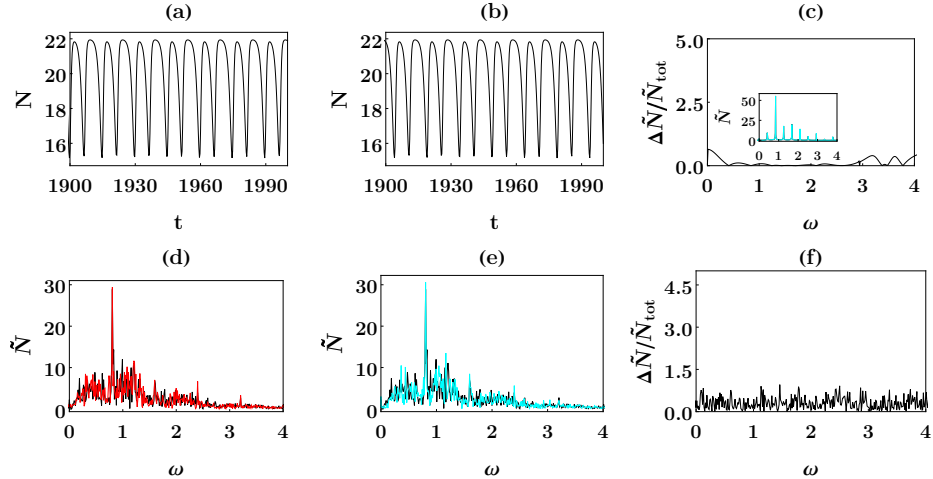}
\caption{{Side-by-side comparison of a time evolution in the
quasi-periodic case ($c=1.799$) obtained by means of (a) numerics,
(b) with the Galerkin approximation. (c) The relative error between
Fourier spectra in the time domain for the total norm presented in
(a) and (b) with inset presenting directly the agreement between
both functions. Bottom row: the behavior of the system in the
chaotic regime at $c=1.81$.  Panel (d) shows a comparison of Fourier
spectra for numerical simulations with two different initial
perturbations; panel (e) an analogical comparison between two
spectra taken from numerical simulations (black), and GA (cyan).
Panel (f)  -- the same as in (c) for the chaotic behavior case.}}
\label{spectra}
\end{figure*}

The difference between GA and GPE numerics becomes non negligible
when the comparison  is done in the chaotic regime (compare
Fig.\ref{spectra}e). Both spectra present qualitatively the same
behavior, but cannot be identical, due to the critical sensitivity
to the initial conditions in the chaotic regime. Notice that the
main peaks obtained using both methods are similar and the
differences appear in the magnitudes of less important peaks.
Interestingly, the same degree of "disagreement" is present when two
numerical simulations, obtained for slightly different initial
perturbations, or more generally for slightly different initial
conditions (cf. Fig.\ref{spectra}d) are compared. In order to better
illustrate discrepancies between both spectra in Fig.~\ref{spectra}f
the relative error is presented in analogy to Fig.~\ref{spectra}c.
The finite numerical accuracy  of GPE solution as well as a
truncation of the expansion  \eqref{Fourier expansion} leads to a
noisy behavior. It can be compared with the quasi-periodic case with
spectrum in Fig.~\ref{behaviour plot}f. As shown in
Fig.~\ref{spectra}d for quasi-periodic case, the relative error is
zero almost in the whole range of the spectrum. That confirms the
very good agreement between GA and numerics in this case (the GA
spectrum is shown  in the inset).

From the above studies we conclude that GA shows practically the
same dynamics as our original system, and all final, asymptotic
states are identical in both cases. Thus the dynamics of two one
dimensional coupled rings may be reduced to a system of coupled
oscillators. This reduction allows to apply GA to obtain
two-dimensional map of final states with varying $c$ and $\gamma$
(the case with varying $\Gamma$ has been studied numerically and
presented in Fig.~\ref{bifurcation diagram}d). From the physical
point of view, the relation between gain and loss ($\gamma/\Gamma$)
is significant. Hence, without a loss of generality, we present an
analysis of solutions in 2D plane $c-\gamma$ for fixed $\Gamma=1$ in
Fig.~\ref{sol_map} (left-top panel). As an input we apply
\eqref{numerical input} with the excitation of all perturbation
modes in a range $l\in\left[-5,5\right]$. It is hence a symmetric
state with an antisymmetric perturbation. In the bottom row of
Fig.~\ref{sol_map} cross-sections of that map for two particular
values $\gamma=2$ and $\gamma=1.75$ are compared. In the top row the
initial perturbation contains excitations of all modes between -5
and 5, while in the middle row the perturbation is magnified 5
times. The bottom row corresponds to the same perturbation amplitude
like in the top row, however, only odd modes in the range
$\left[-5,5\right]$ are excited.

\begin{figure*}[th]
\centering
\includegraphics[scale=0.15]{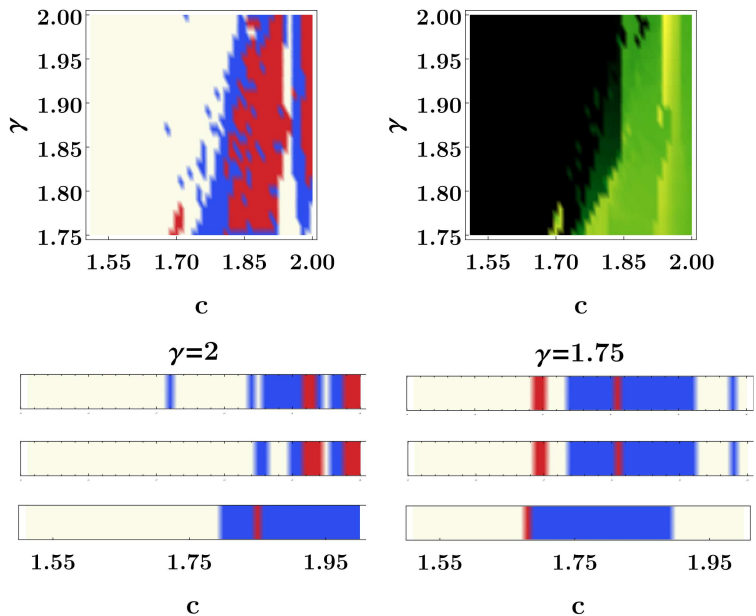}
\caption{Top-left panel: A  map of final states for a symmetric
input with antisymmetric perturbation (excitation of all
perturbation modes between $l=-5$ and $l=5$), Blue (red) color
denotes limit-cycle states (chaotic solutions). In the white area
stationary regular solutions exist only. Top-right panel: the
maximal positive Lyapunov displayed for solutions in the top-left
map. Bottom: cross-sections of the top-left map for $\gamma=2$
(left) and $\gamma=1.75$ (right); each row corresponds to the
different initial perturbation: the top row: all $l=-5$ to $l=5$
modes are perturbed, the middle row: the same for 5 times stronger
perturbation, the bottom row:  only odd  modes in $[-5,5]$ interval
were perturbed.} \label{sol_map}
\end{figure*}

Furthermore, to quantify the chaotic behavior, the maximal positive
Lyapunov exponents have been determined in the same map
($c-\gamma$). The calculation utilizes \eqref{Fourier components
equations} where each amplitude is assumed to take the form

\begin{equation}
B_{\alpha,n}\left(t\right)=e^{-i\mu t}\left[ B_{\alpha s,n}\left(t\right)+ e^{\lambda t}U_{\alpha,n}+e^{\lambda^{\ast} t}V_{\alpha,n}^{\ast} \right], \label{perturbation}
\end{equation}
where index $\alpha=\left\lbrace 1,2\right\rbrace$ pertains to the
ring number, $n$ is an order of the amplitude in the Fourier
expansion \eqref{Fourier expansion} and the subscript $s$ denotes
the exact solution of the system \eqref{Fourier components
equations}. Amplitudes of the perturbation should be small, i.e.
$U_{\alpha,n},V_{\alpha,n} \ll B_{\alpha s,n}$. For simplicity of
calculations we assume that the perturbation depends on time only
via an exponential expression with $\lambda$ being Lyapunov
exponent. Any out-of-equilibrium state (including chaotic one) is
detected by a real, positive value of $\lambda$ and is plotted in
the top-right panel of Fig.~\ref{sol_map}. Generally, the existence
of the area of positive exponents and its location confirms former
results and is in a relatively good agreement with non-stationary
states plotted in the left panel of Fig.\ref{sol_map}.

\section{Conclusions}

The model of two coupled nonlinear Schr\"{o}dinger equations with a
linear gain and a nonlinear loss has been investigated. Such a
system may physically correspond to models of microresonator
nanostructures, polariton condensates or coupled optical waveguides.
Our model may also represent systems with saturation and linear
loss. It seems suitable for description of quantum circulation in a
macroscopic polariton spinor ring condensate \cite{Snoke}. And last
but not least, it can be extended to study the dynamics of more
complicated structures involving inhomogeneous coupling and stripe
of coupled rings or matrix of rings.

We have studied in detail  a family of non-stationary oscillating
states and a transition from regular solutions to chaotic ones
caused by changing one control parameter. On the route to chaos
regular non-stationary states (limit cycles) can be observed,
characterized by a single or multi-frequencies. As we sweep the
value of the coupling from small to large values we encounter many
bifurcation types like saddle-node bifurcation (also in the inverse
realization), as well as period doubling and Hopf bifurcation. All
our numerical results from the "full" model (using equations
(\ref{GPE system})) are quantitatively indistinguishable from that
obtained using a few-modes model - obtained by introducing the
Galerkin Approximation. It allows to limit the number of degrees of
freedom of the system to the most relevant ones and transforms the
problem into a set of nonlinearly coupled modes. This reduction let
us investigate characteristics of the system more effectively and is
the manifestation of the correspondence between the continuous
operator problem and its discrete representation.

We have visualized different routes to chaos in our non-hermitian
system using simple global variables such as the norm of the studied
field $\Psi$ analyzing both the Fourier transform of the norm as
well  as its time delay diagram. Interestingly the transition to
chaos may be also observed in the dynamics of the probability
currents flowing between the studied rings.

\begin{acknowledgements}
We are grateful to Bruno Eckhardt for interesting discussions.
N.V.H. appreciates the hospitality of M.T. during his visit to
Warsaw for scientific cooperation. K.B.Z. acknowledges support from
the National Science Centre (Poland) through project FUGA No.
2016/20/S/ST2/00366. J.Z. acknowledges support by PL-Grid
Infrastructure and EU project the EU H2020-FETPROACT-2014 Project
QUIC No.641122. This research has been also supported by National
Science Centre (Poland) under projects 2016/22/M/ST2/00261 (M.T.)
and 2016/21/B/ST2/01086 (J.Z.). N.V.H. was supported by Vietnam
National Foundation for Science and Technology Development
(NAFOSTED) under grant number 103.01-2017.55.
\\\\
\textbf{Conflicts of Interests:} The authors declare no conflict of
interest.
\end{acknowledgements}

\bibliographystyle{unsrt}


\end{document}